\documentclass[aps,prb,twocolumn]{revtex4-1}

\usepackage{graphicx}
\usepackage{multirow}

\usepackage{amsmath}
\usepackage{color}
\usepackage{ulem}

\begin{document}

\title{Itinerant--electron magnetism: the importance of many-body correlations}

\author{Markus Holzmann}
\affiliation{Univ. Grenoble Alpes, CNRS, LPMMC, 38000 Grenoble, France}
\affiliation{Institut Laue Langevin, BP 156, F-38042 Grenoble Cedex 9, France}

\author{Saverio Moroni}
\affiliation{CNR-IOM DEMOCRITOS, Istituto Officina dei Materiali, and SISSA Scuola Internazionale Superiore di Studi Avanzati, Via Bonomea 265, I-34136 Trieste, Italy}

\begin{abstract}
Do electrons become ferromagnetic just because of their repulsive Coulomb interaction?
Our calculations on the three-dimensional electron gas imply that itinerant ferromagnetism of delocalized electrons without lattice
and band structure, the most basic model considered by Stoner, is suppressed due to many-body
correlations as speculated already by Wigner, and a possible ferromagnetic transition lowering the
density is precluded by the formation of the Wigner crystal. \end{abstract} 
%\pacs{02.70.Ss}

\maketitle

%\section{Introduction}

In 1929, Felix Bloch addressed the possibility of itinerant ferromagnetism 
\cite{Bloch1929} where the same electrons forming the conducting state
give also rise to ferromagnetism.
Considering the free homogeneous electron gas (jellium) as a minimal model
to describe electrons in a metal, he concluded that the exchange energy may lead 
to a ferromagnetic state at densities slightly below those occurring in alkali metals.
Considering correlation between positions of electrons with antiparallel spin,
Wigner \cite{Wigner1934,Wigner1938} approximately calculated the correlation energy -- the gain of energy
compared to the Hartree-Fock approximation -- and pointed out the possibility of crystalline order at low densities. In the same paper \cite{Wigner1934},
he also anticipated
that the magnitude of the correlation energy is important for questions of para-- and
ferromagnetism modifying Bloch's theory on itinerant magnetism.
Later,  Stoner \cite{Stoner1938} predicted the occurrence of a continuous transition
between zero and full magnetization at zero temperature
introducing a repulsive energy term between opposite spin electrons
to phenomenologically account for correlation effects.
The threshold of the
ratio between this repulsive interaction constant and the Fermi energy,
is now commonly known as Stoner criterion.

%\sm{
%\sout{
%In this paper, we show that Stoner's instability is precluded by the transition to the
%Wigner crystal and argue that itinerant magnetism is quite generally suppressed by
%correlation effects in the ground state of homogeneous quantum fluids with 
%spin--independent repulsive interactions.
%}
%}

The question whether pure Coulomb interactions between electrons can drive
a magnetization transition without accompanying structural changes has a long
and controversial history, starting from Bloch's prediction \cite{Bloch1929}.
% in 1929. 
Based on the Hartree-Fock
approximation, he considered the possibility of a first order (discontinuous) transition to
a fully polarized, ferromagnetic electron liquid for electronic densities, $n$,
slightly lower than those of alkaline metals,
$r_s \equiv a/a_B > 5.45$,
where  $a_B$ is the Bohr radius and $a=(4 \pi n/3)^{-1/3}$ is the
mean electron distance.
The first variational Monte Carlo (VMC) calculations \cite{Ceperley78} 
taking into account electron correlations
without relying on perturbative high or low density expansions
indicated a totally polarized quantum fluid for $r_s>26(5)$ before the 
occurrence
of a Wigner crystal at $r_s=67(5)$.
More accurate calculations based on fixed-node diffusion Monte Carlo (DMC) and
transient estimates releasing the nodal constraint subsequently shifted the
ferromagnetic liquid  to lower densities \cite{CeperleyAlder,CAbis}, 
$75(5) < r_s <  100(20)$.
Later calculations \cite{ACP} predicted a partial polarization of 
up to $50 \%$
in the range $20 \lesssim r_s \lesssim 100$.
Two decades later, new calculations \cite{Ortiz1999} with larger system sizes
observed a continuous transition from the paramagnetic to the ferromagnetic
fluid around $r_s = 20(5)$ reaching full polarization around $r_s=40(5)$
before freezing at $r_s = 65(10)$.
%taking better into account electron correlations,
%notably based on quantum Monte Carlo methods,
%have shifted the expected magnetic transition in the three dimensional 
%electron gas model towards significantly lower densities 
%\cite{Ceperley78,CeperleyAlder,CAbis,Ortiz1999}.
The most recent quantum Monte Carlo (QMC) calculations \cite{Zong2002},
reducing systematic errors of the thermodynamic limit extrapolation and
of the fixed-node  bias, 
again support Stoner's picture 
of a continuous magnetic transition, but with an  onset of
partial spin polarization at $r_s = 50(2)$
and completion of full polarization at $r_s \approx 100$, 
just before Wigner crystallization which is
estimated to occur
at $r_s = 106(1)$ in Ref.~\onlinecite{Drummond04}.

Physical realization of such low density electron liquids at low temperatures
is extremely challenging, and experimental findings \cite{Young99} are
controversial
due to finite temperature and band structure effects \cite{Ichimaru,OritizC}.
Recent experimental efforts have been devoted to realize Stoner's model within ultracold atomic gases \cite{Ketterle,Firenze}, where the interaction between two fermions is essentially described by
momentum and energy independent s-wave scattering.
%In an idealized model with strictly contact interaction, the polarized phase
%is actually non--interacting, so that it never crystallizes and 
%
However, there, the strong repulsive s-wave interaction is intrinsically connected with 
a short range interparticle bound state leading to molecule formation.
Although local spin-correlations have been observed, the interpretation of the
experimental observations is not straightforward.

In this paper, we show that Stoner's instability is precluded by the transition to the
Wigner crystal and argue that itinerant magnetism is quite generally suppressed by
correlation effects in the ground state of homogeneous quantum fluids with 
spin--independent repulsive interactions.
Specifically,
we present new results for jellium in three dimensions
based on a sequence of wave functions 
featuring iterative backflow transformations\cite{bfiter1,bfiter2}, within 
the variational and the more accurate fixed-phase diffusion  
Monte Carlo methods\cite{Wagner16,fpa}. 
Zero--variance extrapolation\cite{bfiter1} of the ground state energies 
at finite system size allows us to reliably control the remaining, 
systematic bias of the fixed-phase DMC calculations.
Finite size corrections due to single-particle shell effects \cite{Lin2001}
and two-body terms \cite{fse1,fse2,drummond08} are
applied for thermodynamic limit extrapolation.
Improved accuracy proves crucial, as our calculations show that 
many-body correlations of the ground state wave function favor 
the unpolarized phase of the electron liquid compared to 
partial or fully polarized states and eventually prevent 
itinerant magnetism in jellium
at any densities above crystallization.
We also update the density of the transition to the 
Wigner crystal to a slightly lower value, $r_s=113(2)$.

Methodological and computational improvements of the accuracy 
of QMC calculations thus seem to parallel the
historical developments on the two dimensional electron gas. The
first QMC results \cite{Ceperley78,Rapisarda} indicated an intermediate region
with a fully polarized liquid ground state, shrinking, subsequently, to a weakly first-order
polarization transition occurring just before Wigner crystallization \cite{Varsano,Attac},
and, finally, disappearing completely \cite{Drummond09} following improvements to the fixed node 
bias and finite size methodology.

Taking further into account the absence of a polarized phase in 
$^3$He, a strongly correlated Fermi liquid with an effective dominant hard-core 
interaction, we arrive at the quite general conclusion
that pure correlation effects in homogeneous quantum liquids tend to
suppress itinerant
magnetization, in contrast to common argumentations based on Stoner's model.
In these systems,
crystallization, the
spontaneous breaking of translational symmetry, seems to win the competition
against spontaneous magnetization.

%\section{Methods and Results}

In the following, we describe the details of 
our numerical methods to determine the 
low-density ground state phase diagram of 
jellium -- non-relativistic electrons interacting via Coulomb's potential
with each other and with a homogeneous positive background to guarantee charge
neutrality \cite{vignale,book}. The ground state energy per electron
of the model at three values of the electronic density, $n$, corresponding 
to $r_s=70$, $100$, and $120$, and six different spin polarizations
$\zeta=0.0$, 0.18, 0.42, 0.61, 0.79, and 1.0,
is addressed by variational and diffusion Monte Carlo 
simulations \cite{Wagner16} of a finite system containing $N=66$ electrons 
imposing periodic boundary conditions
for the particles' positions;
the long-range Coulomb potential is evaluated by standard splitting 
into real and reciprocal space contributions \cite{Ewald,Natoli}. 
\begin{figure}[h]
\includegraphics[width=9cm]{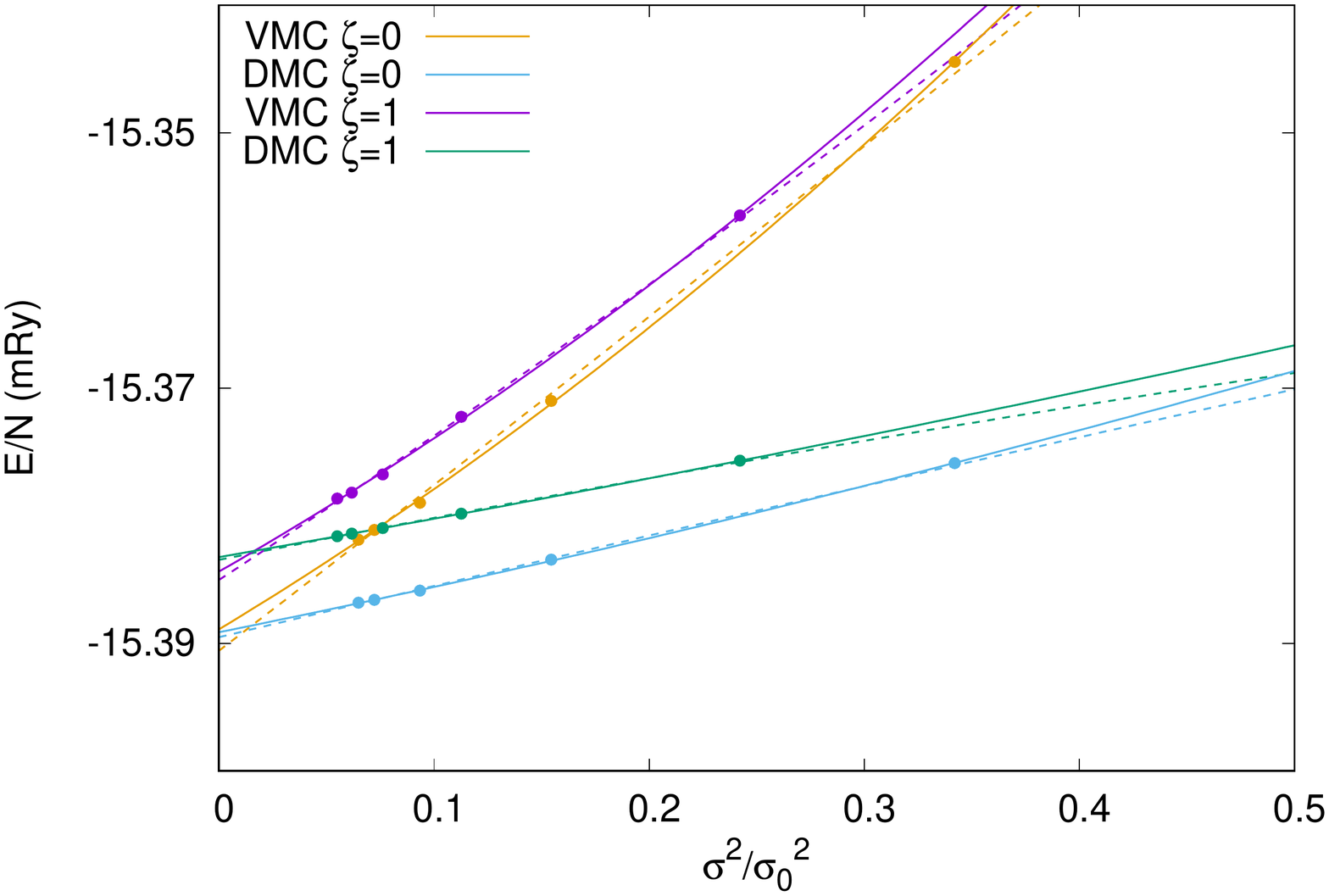}
\caption{
Extrapolation of the energy to zero variance for $r_s=100$ at 
polarizations $\zeta$=0 and 1. The data are calculated with with VMC and DMC 
using SJ, BF0, $\ldots$, BF4 wave functions in order of decreasing energy.
The reference value $\sigma_0^2$ is the variance of the local energy at $\zeta$=0 with the SJ wave function.  
The curves are quadratic fits; for each set of data points (VMC and DMC 
for $\zeta$= 0 and 1) there are
two curves, one of which (solid line) excludes the SJ energy from the fit.
}
\label{fig_eofsigma}
\end{figure}
\begin{figure}[h]
\includegraphics[width=9cm]{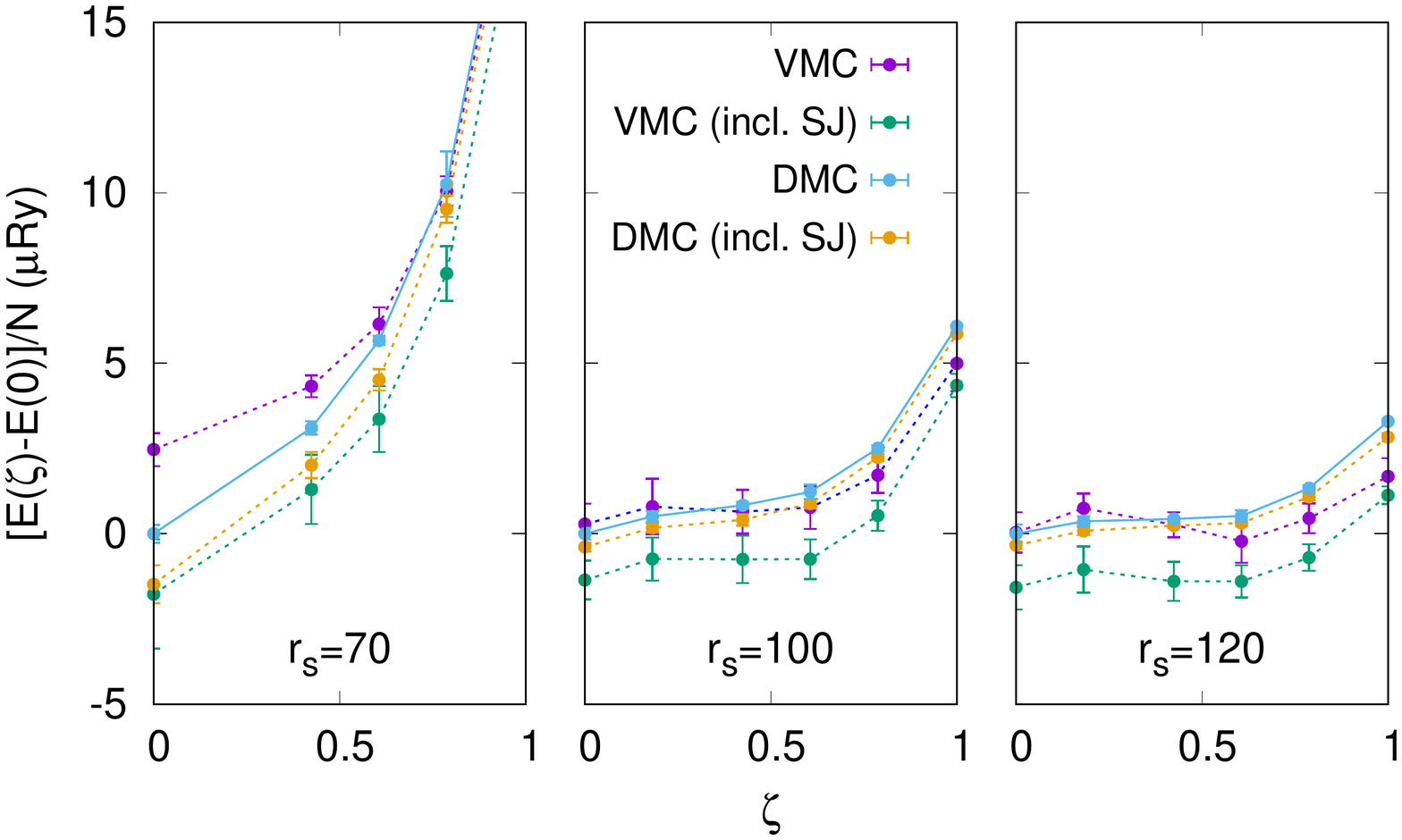}
\caption{
The polarization energy $E(\zeta)-E(0)$ obtained from zero--variance extrapolations of the VMC and DMC
energies, 
including or excluding the SJ result.
%with or without the SJ result. 
The common reference energy $E(0)$ is 
the zero--variance 
extrapolation of the DMC energy 
without 
%including 
the SJ data.
%excluding the SJ result.
%extrapolation without the SJ result of the DMC value.% $E(\zeta=0)$.
}
\label{fig_fitout}
\end{figure}

In the DMC runs\cite{supp}, the number of walkers is 1280 and the time step is 15, 20, and
30 Ry$^{-1}$ for $r_s=70$, 100, and 120, respectively. The estimated time step 
error is 10$^{-7}$ Ry or less, which is about the size of the statistical 
error on our final results (the zero--variance extrapolation of the DMC 
energy, see below). The population control bias is even smaller, 
of the order of 10$^{-8}$ Ry.

The accuracy of the ground state energy of a finite system
is limited by the underlying many-body trial wave function $\Psi$ used 
for calculating expectation values in VMC \cite{Wagner16} and for imposing 
the phase in DMC \cite{fpa}, respectively. In order to remove such a bias,
we consider a series of trial wave functions
of increasing quality, starting from the standard Jastrow--Slater
and backflow forms (SJ and BF0)\cite{Zong2002}, and 
adding up to four iterative backflow
transformations (BF1,$\ldots$,BF4)\cite{bfiter1}.

Specifically,
the Jastrow-Slater wave function SJ({\bf R}) 
explicitly depends on the coordinates ${\bf R}$ of all the particles
through the two--body pseudopotential $u_0$
%and three--body pseudopotentials $u_0$ and $\xi_0$ 
in the Jastrow factor $\exp(-U_0({\bf R}))$, and through plane--wave orbitals 
in the Slater determinant $D({\bf R})$.
We then recursively build sets of transformed coordinates 
${\bf Q}_0,\ldots,{\bf Q}_k$, where ${\bf Q}_i$ depends on ${\bf Q}_{i-1}$
through the $i$--th backflow pseudopotential 
$\eta_i$, with ${\bf Q}_{-1}\equiv{\bf R}$. The $k$--th iterative backflow wave function is
\begin{equation}
{\rm BF}k({\bf R})=\exp(-U_0({\bf Q}_{-1})-\ldots-U_k({\bf Q}_{k-1})
)D({\bf Q}_k).
\end{equation}
For the backflow wave functions we include both two-- and three--body 
pseudopotentials $u_0$ and $\xi_0$ in $\exp(-U_0)$, and only two--body 
pseudopotentials $u_i$ in $\exp(-U_i)$ for $i=1,\ldots,k$.
The plane--wave orbitals in the Slater determinant are evaluated at the
last set of transformed coordinates, ${\bf Q}_k$.

The two-body pseudopotential $u_0$ is initially of the the RPA form 
\cite{Ceperley78} with an explicit long-range part in Fourier space 
spanning the first 20 shells of reciprocal vectors, and
the real-space part represented by locally piecewise-quintic
Hermite interpolants (LPQHI) with 8 degrees of freedom which are 
subsequently treated as optimization parameters.
The three-body pseudopotential ($\xi_0$), the backflow pseudopotentials 
($\eta_i$ with $i=0,\ldots,k$), and the two--body pseudopotentials in 
the transformed coordinates ($u_i$ with $i=1,\ldots,k$) are all expressed as  
LPQHI with 6 degrees of freedom each, with the exception of $\eta_0$ which
is augmented with 5 shells of Fourier components.
The LPQHI coefficients of all the pseudopotentials, as well as
the Fourier components of $\eta_0$, are optimized 
independently for each wave function in the hierarchy.

The energy $E$ computed for $r_s=100$ in VMC and DMC simulations 
using all the above wave functions is plotted in Fig.~\ref{fig_eofsigma} 
against the corresponding VMC variance 
$\sigma^2=\langle\Psi|(H-\langle\Psi|H|\Psi\rangle)^2|\Psi\rangle$. 
The exact ground--state
energy, which has zero variance, can be reliably estimated by extrapolation
\cite{bfiter1}, given the smoothness of the data over a significant range 
extending to very low values of $\sigma^2$.
We assume a quadratic dependence of $E$ on $\sigma^2$. 
Since the range of validity of such a dependence is not known, we perform
the extrapolation with and without the highest 
energies and variances, obtained with the SJ wave function. The result does 
not change significantly if we include the SJ result and/or switch between VMC 
and DMC data for the extrapolation. In particular,
Figure \ref{fig_fitout} shows that the polarization energy is only marginally
influenced by the choice of the data set. 

%Twist-averaged boundary conditions of the trial wave function \cite{Lin2001}
%on a regular grid of $1000$ twist angles are used to reduce shell effects 
Twist-averaged boundary conditions \cite{Lin2001,supp}
are used to reduce shell effects 
of the finite simulation cell and afford thermodynamic limit 
extrapolation without resorting to large simulation cells. 
Residual single particle shell effects due to the 
discrete twist grid and to reduced-symmetry open-shell fillings for finite
polarizations with $N=66$, $\Delta T_0$, are estimated from the 
non-interacting electron gas. Two--particle finite size corrections
for the potential and kinetic energy, $\Delta FSE$, are addressed by 
interpolation of the long--range part of the static structure factor 
and by the analytical long--range expressions for the two--body and
backflow pseudopotentials $u_0$ and $\eta_0$ of the wavefunction 
\cite{BF03,fse_rpa,fse2}.
In contrast to the high density limit \cite{Ruggeri18,Spink13}, where
residual size effects in the kinetic energy introduce quasi-random fluctuations
in the extrapolation, 
these effects are suppressed in the large $r_s$ region addressed here
(see Supplementary Material \cite{supp}). Furthermore, at low densities, 
the corrections $\Delta FSE$ are largely dominated 
by the zero--point energy of the plasmon \cite{fse1}.
Whereas the single particle size corrections depend on the spin polarization, 
the long--range structure factor does not reveal any
systematic dependence on $\zeta$ within the statistical error of the
present simulations. 
Therefore, we average the structure factor over spin-polarizations 
in the calculation of $\Delta FSE$, so that the final polarization 
energy is not affected by statistical fluctuations in the estimates 
of $\Delta FSE$. 
Only the absolute value of the estimated ground state energy, used below to
locate the Wigner crystallization, is 
then susceptible to the details of the calculation of $\Delta FSE$. 

The results for the energies and the variances obtained with different 
trial wave functions, the zero--variance extrapolations, and
the finite--size corrections are collected in the Supplemental Material \cite{supp}.
Note that the variance extrapolation is done on the energies of the 
finite--size system, and size corrections (for the polarization energy 
and the Wigner crystallization) are applied afterwards,
using the value of $\Delta FSE$ obtained at the highest wave function level.
%are summarized in Tables \ref{table_rs70} to \ref{table_rs120}.  
\begin{figure}[t]
\includegraphics[width=9cm]{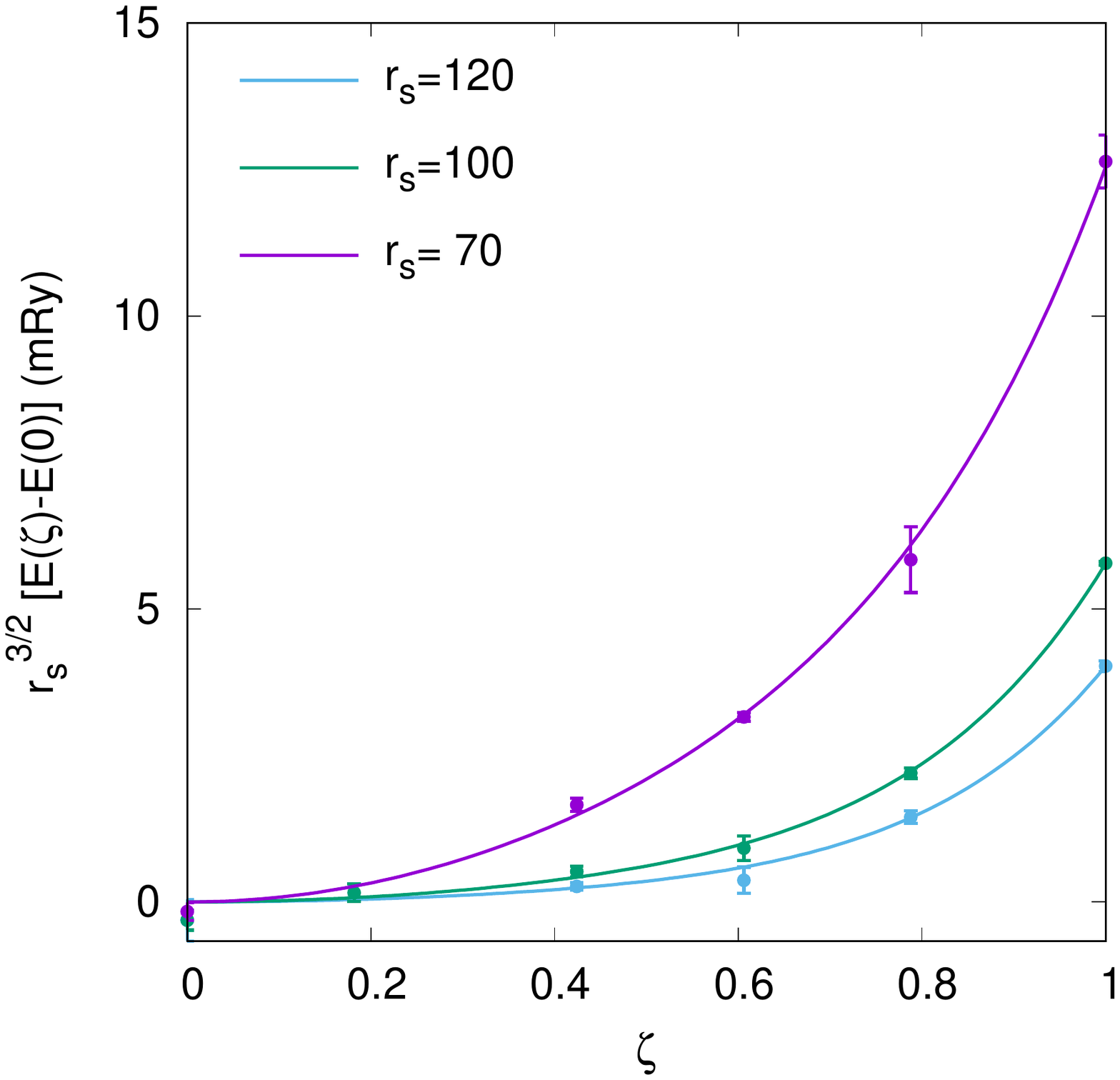}
\caption{
The polarization energy $E(\zeta)-E(0)$ obtained from zero--variance 
extrapolations of the DMC energies without the SJ result. The lines are 
polynomial fits with terms of order 0, 2 and 6. 
Alternative functional forms and ensuing confidence levels for 
our conclusions are discussed in the Supplemental Material\cite{supp}.
}
\label{fig_eofzeta2}
\end{figure}
\begin{figure}[h]
%\begin{figure}
\includegraphics[width=9cm]{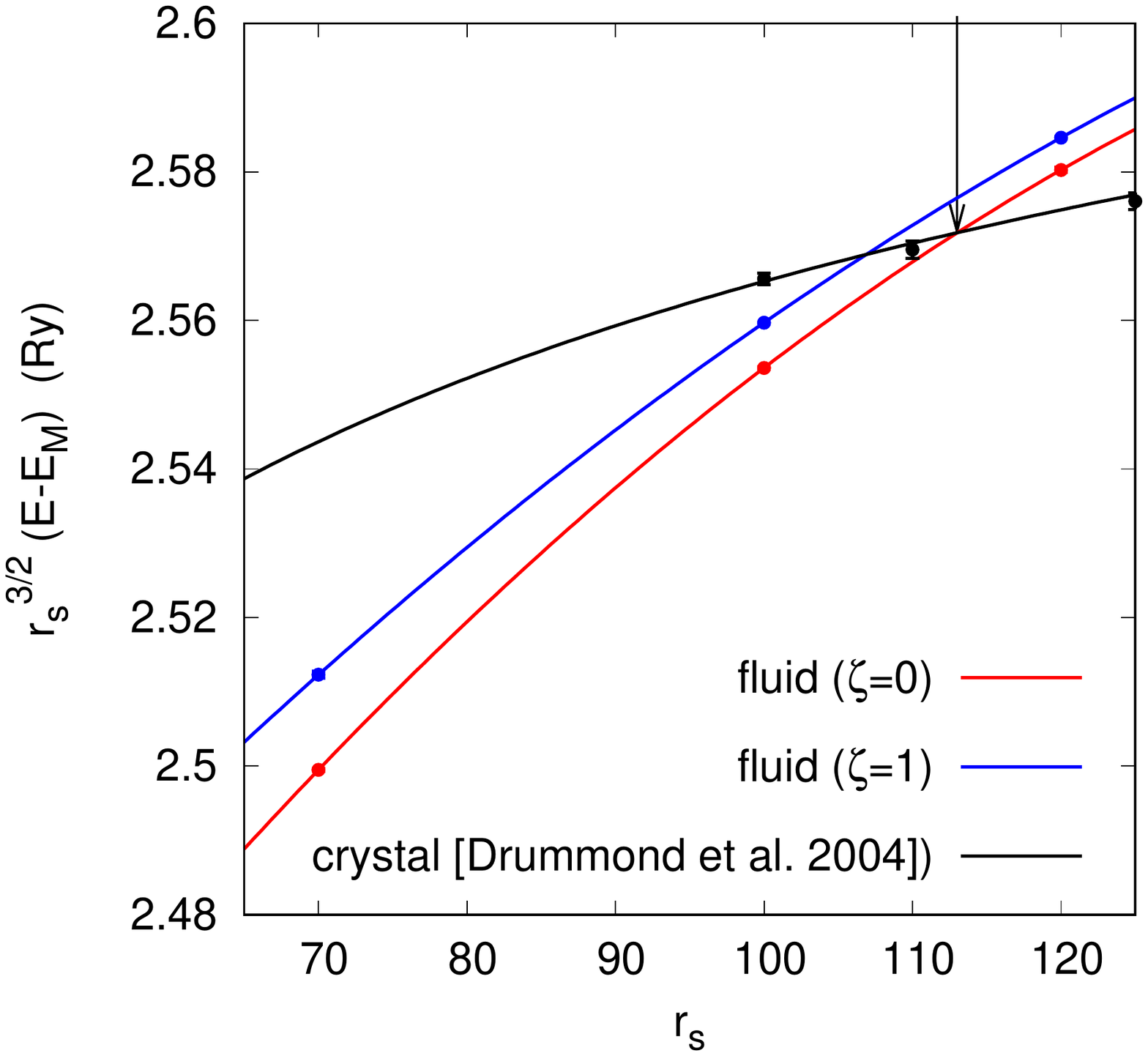}
\caption{
DMC energy as a function of $r_s$ for the paramagnetic and the 
ferromagnetic fluid and for the Wigner crystal. 
$E_M=-1.79186/r_s$ Ry is the Madelung energy of the bcc 
lattice\cite{Drummond04}. The lines are quadratic interpolations for
the fluid phases, and a fit of the form\cite{Drummond04} $a+b/\sqrt{r_s}$ for the Wigner
crystal.
The arrow at $r_s=113$ locates the crystallization point.
}
\label{fig_wigner}
\end{figure}

The final polarization energy of jellium at low densities, our main 
result, is shown in Fig. \ref{fig_eofzeta2} for $r_s=$70, 100 and 120. 
It is obtained from the zero--variance extrapolation of the DMC energy, 
excluding the SJ result. This choice gives the smoothest polarization 
energy, as well as the lowest $\chi^2$ in the fit to 
the energy vs. variance data, but it is otherwise uninfluential for the
%conclusion that $E(\zeta)$ is higher than $E(0)$ for all the densities
%considered, and therefore a partially or fully polarized
%state is never stable.\cite{supp}
oucome that $E(\zeta)$ is higher than $E(0)$ for all the densities
considered, and therefore a partially or fully polarized
state is never stable. Confidence levels for our conclusion are 
given in the Supplemental Material\cite{supp}.
%Note that the variance extrapolation is done on the energies of the 
%finite--size system, and size corrections (for the polarization energy 
%and the Wigner crystallization) are applied afterwards.

The zero--variance extrapolation of the DMC energy, corrected for 
finite--size effects, is compared in Fig. \ref{fig_wigner} with the 
fixed--node DMC energy \cite{Wagner16} of the Wigner crystal of 
Ref. \cite{Drummond04} as a function of $r_s$. For the crystal phase, 
finite--size effects are assessed using large simulation cells with 
up to 512 electrons \cite{Drummond04}. This procedure differs
from that used in the present work for the liquid phase, but it should be
equally reliable. The fixed--node DMC bias\cite{Wagner16} for the crystal phase 
is negligible: it is bounded by (and presumably much smaller than)
the difference between the fixed--node energy
and the exact bosonic ground--state energy, which we find to be of the
order of the statistical error on the crystal data of Fig. \ref{fig_wigner}. 
The critical value for the Wigner crystallization is shifted 
to $r_s=113(2)$, where the uncertainty includes both the statistical error
of the simulation data and a conservative estimate of the residual bias in the
size corrections\cite{supp}.

%\section{Conclusions}

In this paper, we have presented accurate quantum Monte Carlo
calculations addressing the possibility of a magnetically
polarized fluid in the ground state phase diagram of the homogeneous 
electron gas.  We have shown that iterated backflow wave 
functions \cite{bfiter1,bfiter2} provide highly accurate
results for the energy and very low values of its variance, such that a zero 
variance extrapolation provides fairly unbiased results for 
the polarization energy. Our calculations clearly demonstrate that the 
simple mean-field picture based on Stoner's model
is not sufficient to explain itinerant ferromagnetism
as the partially or fully polarized fluid state is unstable
versus Wigner crystallization.

Therefore, in addition to repulsive interparticle interactions,
band structure effects must play an essential role for the
occurrence of itinerant ferromagnetism in real materials.

Similar results have been found for liquid $^3$He in two \cite{Nava12,bfiter1} and three dimensions\cite{he,bfiter2},
the two dimensional electron gas \cite{Drummond09}, and two dimensional quantum gases  with repulsive 
dipolar interaction \cite{dipol}, where accurate, quantitative treatment of correlation 
effects have always stabilized the spin-unpolarized phase.

From a more general point of view, Stoner's instability constitutes a reconstruction
of the Fermi surface of the unpolarized to the polarized gas due to interactions.
However, this instability is
quite naturally in competition with the reconstruction of the Fermi surface related to 
spin and charge density waves \cite{Overhauser,HF,Lewin} (not addressed in this work) or Brillouin zone formation
for Wigner crystallization. 
Despite the quite different interparticle interaction, hard or soft core potentials, the
Stoner transition to a spin-polarized phase predicted within mean-field arguments
seems to be quite generally preceded by transition to a crystalline phase for
homogeneous systems with spin-independent interactions.

%\section*{Acknowledgments}
We acknowledge the CINECA award under the ISCRA initiative
for the availability of high performance computing resources and 
the Fondation NanoSciences (Grenoble) 
for support. Part of the computations were performed using 
the Froggy platform of the CIMENT infrastructure, which is 
supported by the Rh{\^o}ne-Alpes region (grant CPER07-13 CIRA)
and the project Equip@Meso (ANR-10-EQPX-29-01) of the ANR.

\widetext
\section*{Supplemental material}
This supplemental materials contains (a) details of the DMC simulations,
(b) a discussion of the interpolation for the polarization energy,
(c) a test of the size extrapolation,
and (d) tables with all the finite--size energies 
and variances for different wave functions, densities and polarizations,
as well as their zero--variance extrapolations and 
finite--size corrections.

\subsection*{Details of the simulations}
\begin{figure}[h]
\includegraphics[width=10cm]{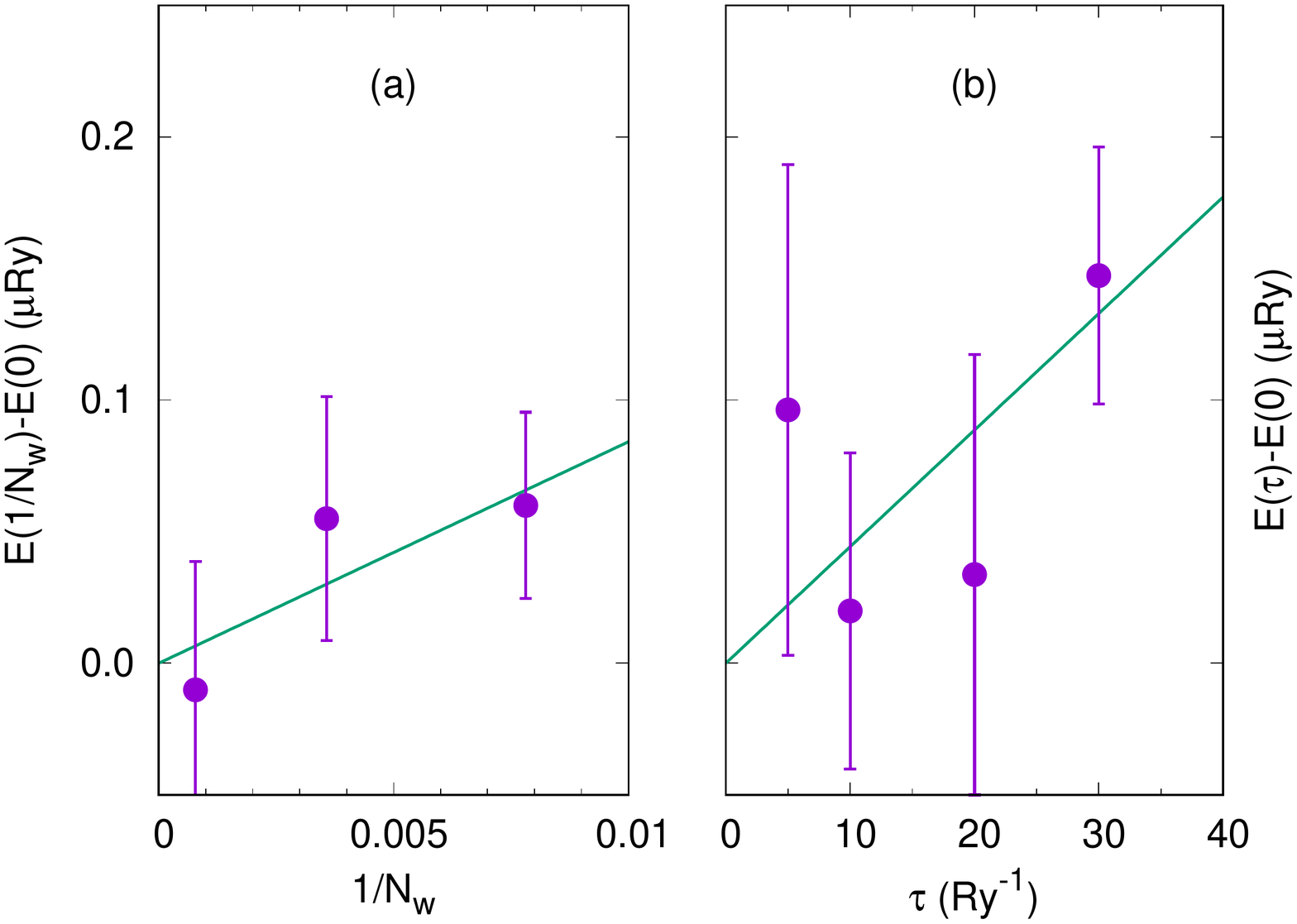}
\caption{
Extrapolation of the DMC energy to (a) infinite number of walkers $N_W$ and
(b) zero time step $\tau$, for $r_s=100$ at polarization $\zeta=0$ using a BF2 trial wave function.
The bias for given $\tau$ or $N_W$ is estimated as the corresponding value of the linear fit.}
\label{extr}
\end{figure}

The calculations presented in the main text have been done for $N=66$ 
electrons in twist averaged boundary conditions, using a regular grid of $1000$
twists $\theta_{\alpha}=(m+1/2) \pi/ML$ with $m=-M,-M+1,\dots,M-1$ and $M=5$
($\alpha=x,y,z$ and $L$ is the linear extension of the cubic simulation cell).
Care must be exerted when employing a regular grid of twists\cite{Ruggeri18}.
Using VMC simulations with a SJ wave function, we have verified that the
twist-averaged energy is converged within the accuracy of our calculations
with respect to $M$, by comparison with a twist mesh with $M=10$ and 
$\theta_{\alpha}=m\pi/ML$ with $m=-M,\dots,M-1$.
We further ruled out any significant bias in our regular--mesh results 
performing tests against unbiased averages over random twists uniformly sampled in 
the Brillouin zone, as well as against exact twist averaging\cite{fse1,fse2,Ruggeri18}; 
some of the results of exact twist averaging are shown in the section on the size extrapolation, see below.
The wave functions are optimized by minimizing the twist-averaged energy.
DMC energies are calculated with 1280 walkers and time step $15$, $20$, and $30$ Ry$^{-1}$
for $r_s=70$, 100, and 120, respectively. Population control bias and time step error are 
estimated to be of order of 10$^{-7}$ Ry and 10$^{-8}$ Ry, respectively, as illustrated in 
Fig.~\ref{extr} for $r_s=100$ at polarization $\zeta=0$ with a BF2 trial wave function.

\subsection*{Interpolation of different functional forms for the polarization energy}

All of our QMC calculations are done on a grid of few different polarizations,
$\zeta=$0, 0.18, 0.42, 0.61, 0.79, and 1, 
yielding a minimum of the energies at the unpolarized state.
Since the polarization energies become less stiff at lower densities, the energy resolution needed
to directly exclude a partially polarized ground state of small polarizations becomes rapidly
unaffordable. The occurrence or absence of a partially polarized state can still be addressed 
considering possible functional forms of the polarization energy, 
$E_p(\zeta) \equiv E(\zeta)-E(0)$.

\begin{figure}[h]
\includegraphics[width=10cm]{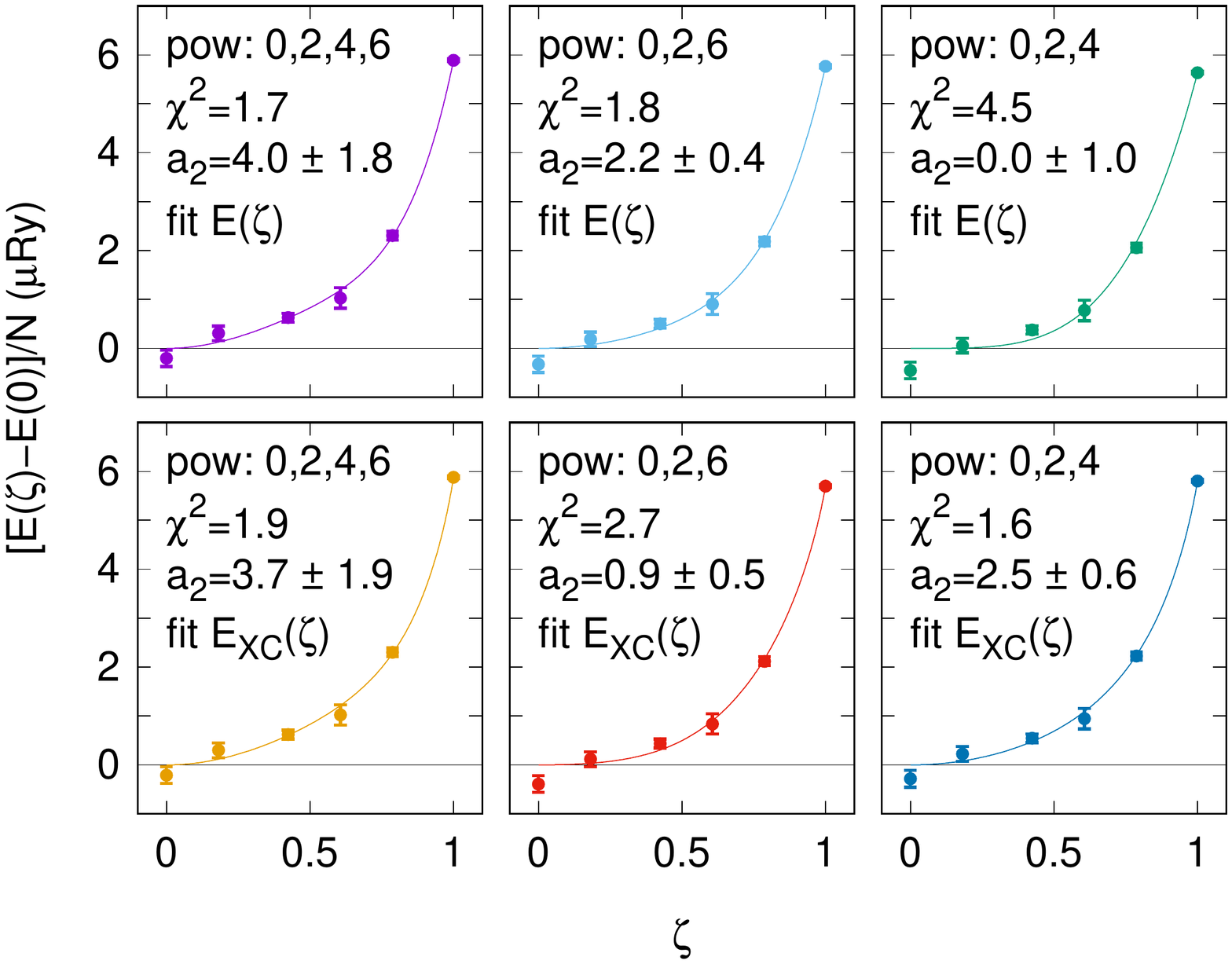}
\caption{
The polarization energy per particle for $r_s=100$ obtained from the best fit of various
functional forms: upper panels, fit of $E(\zeta)$; lower panels, fit of $E_{XC}(\zeta)$; 
left panels, powers 0, 2, 4, 6 of $\zeta$ (see text); middle panels, powers 0, 2, 6; 
right panels, powers 0, 2,4. The reference value $E(0)$ is the energy at zero 
polarization resulting from the fit. The data used are DMC zero-variance 
extrapolations w/o SJ. In each panel, we list the powers of
$\zeta$ employed in the fit, the reduced $\chi^2$, and the resulting
coefficient $a_2$ of the second order term in $E(\zeta)$.
}
\label{fig:fit_rs100}
\end{figure}

\begin{figure}[h]
\includegraphics[width=10cm]{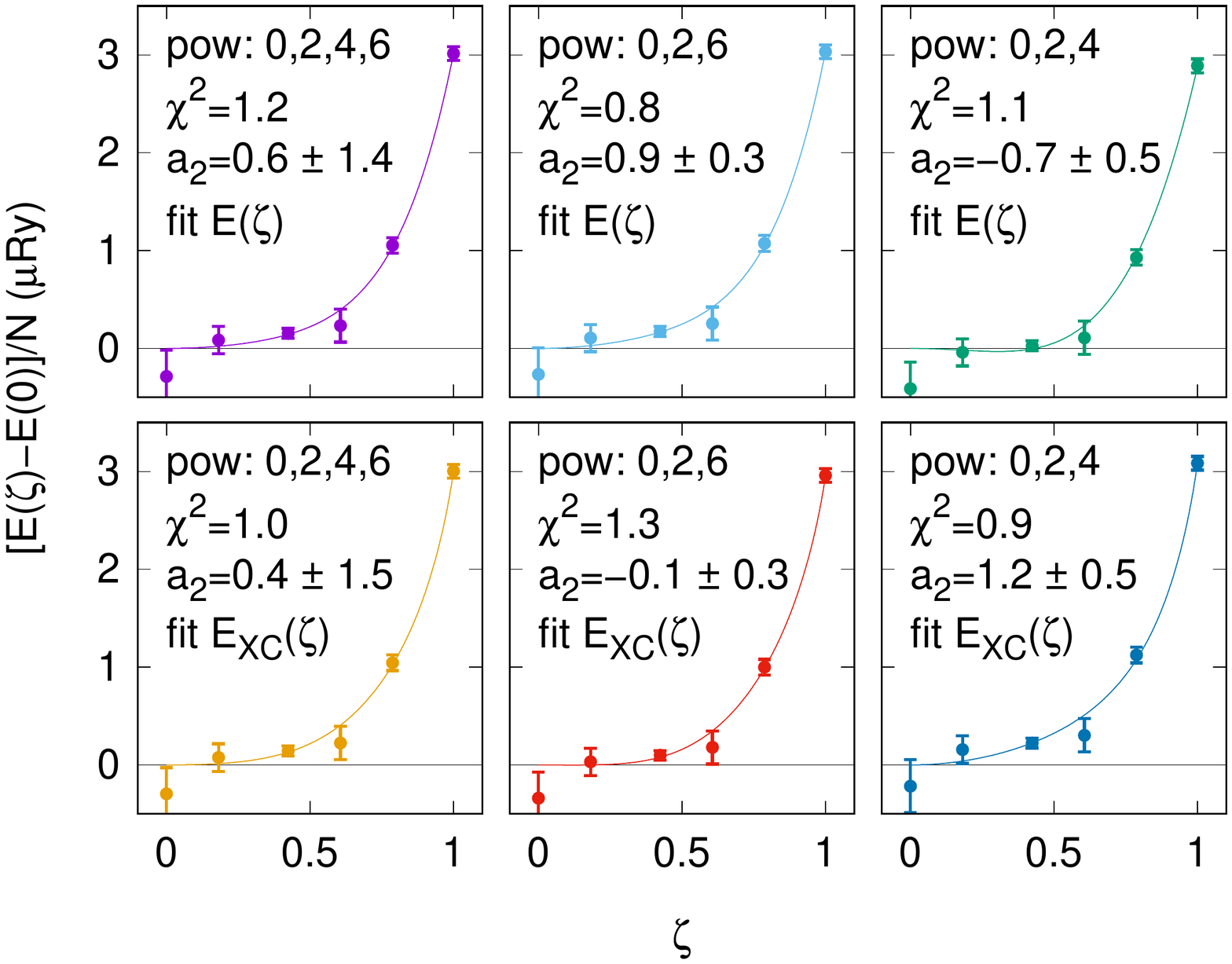}
\caption{
Same as Fig. \ref{fig:fit_rs100} for $r_s=120$.
}
\label{fig:fit_rs120}
\end{figure}

The simplest functional form assumes a polynomial dependence, $E(\zeta)=\alpha_0 +
\sum_n \alpha_n \zeta^n$,
containing only even powers of $\zeta$. 
Given the limited number of data points we can fit, and their statistical errors,
we can only pin down the value of a restricted number of parameters.
In order to gain some confidence on the robustness of the results, we also fit a
completely different
functional form, suggested by the presence of a Slater determinant of plane waves 
in the wave function (either SJ or BF$k$), which assumes a polynomial dependence of the 
exchange-correlation energy, $E_{XC}(\zeta)\equiv E(\zeta)-E_{id}(\zeta)=\beta_0 +
\sum_n \beta_n \zeta^n$, where 
$E_{id}(\zeta)=\left(\frac{9 \pi}{4}\right)^{2/3}\frac{3}{10 r_s^2} 
\left[ (1+\zeta)^{5/3}+(1-\zeta)^{5/3}\right]$ is the ideal gas energy. 
The best fits obtained using both functional forms with various combinations of powers
of $\zeta$ up to 6 are illustrated in Fig.~\ref{fig:fit_rs100} for $r_s=100$ and 
Fig.~\ref{fig:fit_rs120} for $r_s=120$.
For $r_s=100$ all fits give the lowest energy for $\zeta=0$. The 
second order term in the expansion of $E(\zeta)$, related to the inverse spin susceptibility, 
is larger than the statistical uncertainty in all cases but the top right panel, 
which however has a poor $\chi^2$. The second smallest term of second order is found 
in the bottom middle panel, again with a rather large $\chi^2$. 
The inclusion of more powers of $\zeta$ (left panels) begins
to overfit the data, because it does not decrease significantly the $\chi^2$.
Interestingly, however, it does support the fits with fewer parameters which yield
smaller $\chi^2$ and larger second order terms.
For $r_s=120$, $E(\zeta)$ has a minimum for a finite polarization in 
the top right panel (barely visible in the figure) and in the bottom middle panels
(ten times shallower). However the top middle and bottom right panels both give the
energy minimum at $\zeta=0$, with $a_2$ positive and larger in modulus, and smaller $\chi^2$.
Again, the inclusion of more powers of $\zeta$ in the left panels results in a little overfit,
supporting however the fits with fewer parameters which yield smaller $\chi^2$ and positive
second order terms. We conclude that the fluid phase is definitely paramagnetic a $r_s=100$. 
For $r_s=120$ it is most likely paramagnetic, possibly on the verge of spin polarization;
however at this density the ground state is already the Wigner crystal.
It turns out that in all cases $E(\zeta)$, beyond the term quadratic in $\zeta$, is well represented
by only one dominant higher-order contribution, with a positive coefficient. In this situation,
the existence of a partially polarized phase is linked to the sign of $a_2$, and its confidence levels 
can be directly inferred by the values of $a_2$ and their uncertainties, shown in Figs.~\ref{fig:fit_rs100} and \ref{fig:fit_rs120}. 
\begin{table}[h]
\begin{tabular}{|c|c|c|c|}
\hline
data set   & powers of $\zeta$ & $\chi^2$ & $a_2$          \\
\hline
\multirow{2}{*}{VMC full} 
           & 0,6               & 0.2      &  0             \\
           & 0,2,6             & 0.2      &  0.7$\pm$0.6   \\
\hline
\multirow{2}{*}{VMC w/o SJ}
           & 0,6               & 0.1      &  0             \\
           & 0,2,6             & 0.1      &  0.4$\pm$0.5   \\
\hline
\multirow{3}{*}{DMC full} 
           & 0,2,4             & 3.8      &  0.6$\pm$0.9   \\
           & 0,2,6             & 2.4      &  2.5$\pm$0.5   \\
           & 0,2,4,6           & 3.5      &  3.4$\pm$2.6   \\
\hline
\multirow{3}{*}{DMC w/o SJ}
           & 0,2,4             & 4.5      &  0.0$\pm$1.0   \\
           & 0,2,6             & 1.8      &  2.2$\pm$0.4   \\
           & 0,2,4,6           & 1.7      &  4.0$\pm$1.8   \\
\hline
\end{tabular}
\caption{
Reduced $\chi^2$ and coefficient $a_2$ of the second order term in a polynomial fit
of $E(\zeta)$ for $r_s=100$, using various powers of $\zeta$ and various 
data sets.
}
\label{tab:fit_rs100}
\end{table}

We now address the dependence on the choice of the QMC data set. Above,
we have considered  DMC zero-variance extrapolations without the SJ results.
Using the DMC data set which includes the SJ results,
we find generally higher values of $a_2$, albeit with a somewhat larger $\chi^2$.
This makes a stronger case for the paramagnetic fluid.
An example for the fit of $E(\zeta)$ at $r_s=100$ is given in Table \ref{tab:fit_rs100}.
The VMC data sets, either with or without the SJ results, have rather large statistical
uncertainties, so that $E(\zeta)$ is already well fitted using only two powers of $\zeta$.
For both $r_s=100$ and 120 the choice of powers 0 and 6 gives the lowest 
$\chi^2$, and of course $a_2=0$. The inclusion of a quadratic term does not change
the reduced $\chi^2$ and leaves $a_2$ largely undetermined.
For $r_s=100$, we report the results in Table \ref{tab:fit_rs100}.
The VMC results are consistent with a paramagnetic ground state of the fluid phase.

\subsection*{Size Extrapolation: Numerical study with SJ and BF VMC}

Size extrapolations to the thermodynamic limit presents one of the major sources of systematic
bias of our QMC calculations.
By construction, the functional form of backflow wave functions is size consistent 
due to the
extensitivity of the logarithm of the trial wave function, which has been shown explicitly
for Slater Jastrow wave functions with and without backflow \cite{fse1,fse2,drummond08} as well as
with iterated backflow wave functions \cite{bfiter1}. However, explicit numerical extrapolation
to the thermodynamic limit is computational expensive, e.g. as the quality of the optimization
within VMC as well as time step and population bias in DMC must be controlled carefully to
avoid loss of quality in the description for increasing system sizes biasing the extrapolation.
Above and in the main paper, we have used
thermodynamic limit extrapolations based on finite size corrections obtained from correlation
functions \cite{fse2} which have been shown to reach accuracies comparable to
explicit numerical extrapolations.

Typically, reaching the asymptotic region where finite size corrections can be successfully 
applied is easier in the low density region where the potential energy dominates. Quasirandom
contributions, posing difficulties in the high density region $r_s  \lesssim 1$ addressed in Refs.~\onlinecite{Ruggeri18,Spink13},
are expected to be less relevant there. Ref.~\onlinecite{Zong2002} has already shown that twist average,
together with the leading order $1/N$ corrections, provides accurate extrapolations to the thermodynamic limit in the region $50 \le r_s \le 100$. In the following, we provide further tests 
using Slater Jastrow wave functions with the analytical Jastrow potential of Ref.~\onlinecite{BF03}
to verify our finite-size extrapolations. We assume
that size effects are to a large extent transferable to the more correlated wave functions, e.g.
two-particle size corrections from backflow as discussed in Ref.~\onlinecite{fse2} are suppressed by $1/r_s$
and negligible in the considered large $r_s$ region. The SJ VMC study  thus provides an 
independent test of the accuracy of our extrapolations. Transferability has been checked 
in a few relevant cases
using the analytical backflow wave function of Ref.~\onlinecite{BF03}.

\begin{figure}[h]
\includegraphics[width=8cm]{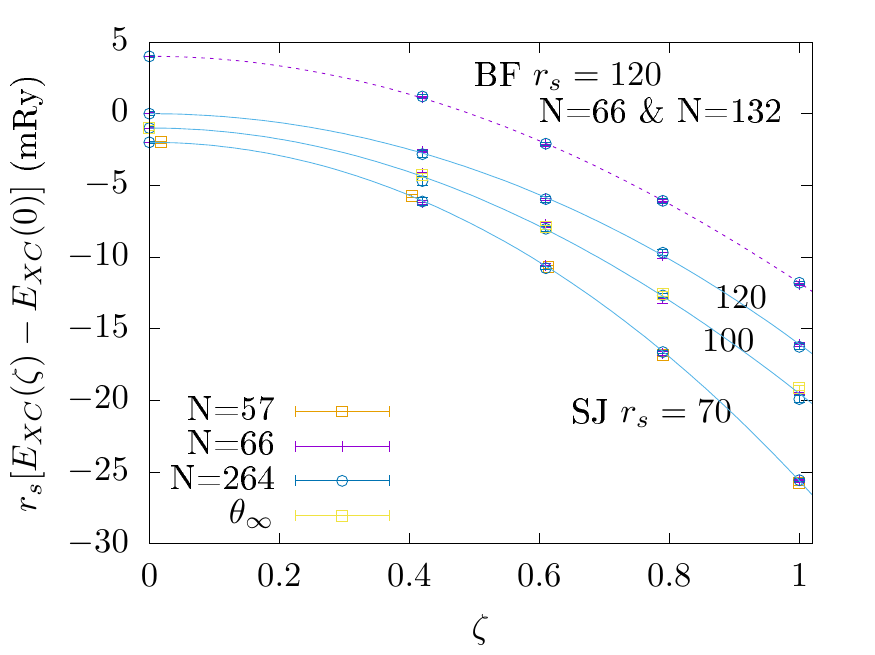}
\caption{
Exchange-correlation part of twist averaged polarization energies, $E_{XC}(\zeta)-E_{XC}(0)$,
for SJ VMC wave functions with analytical Jastrow potentials
for $N=66$ and $N=264$ electrons at the three different densities
together with the corresponding polynomial approximation of degree 2 and 4,
yielding values of $\chi^2$ per degree of freedom of $1.0$ ($r_s=70$), $0.5$ ($r_s=100$), 
and $0.8$ ($r_s=120$). Curves 
for $r_s=100$ and $r_s=70$ are shifted by a negative constant offset.
For $r_s=70$ we plot also the twist averaged energies for $N=57$ electrons  which would correspond
to a closed shell of the fully polarized system at the Gamma point (zero twist). 
With a positive constant offset we plot
the polarization energies of $N=66$ compared to $N=132$ electrons using analytical
backflow wave functions at $r_s=120$. The polynomial fit of degree 2 and 4, shown as dashed line,
has a $\chi^2=0.3$ per degree of freedom.
For $N=66$ at $r_s=100$ we also show, in yellow squares, the results obtained with exact twist
averaging\cite{fse1}.
}
\label{Extrapol_N_SJBF}
\end{figure}
\begin{figure}[h]
\includegraphics[width=8cm]{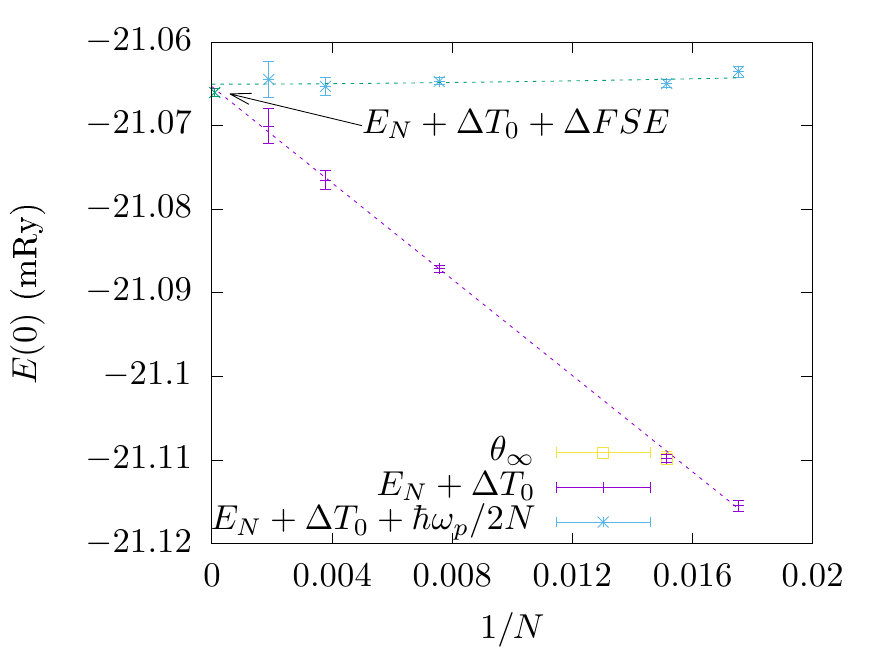}
\caption{
Numerical extrapolation of twist-averaged energies of 
SJ VMC wave functions with analytical Jastrow potential
for $N=57$, $N=66$, $132$, and $264$ electrons at $r_s=70$ and zero polarization
corrected by the ideal gas kinetic energy corrections, $E_N+\Delta T_0$, 
together with
the energy values including the analytical leading order 1/N (Plasmon) correction,
$E_N+\Delta T_0 + \hbar \omega_p/2N$ with
$\hbar \omega_p/2=\sqrt{3/r_s^3}$ Ry.
The lines are based on fitting linear and $N^{-5/3}$ corrections for $E_N+\Delta T_0$
and only $N^{-5/3}$ corrections for $E_N+\Delta T_0 + \hbar \omega_p/2N$.
At the origin we show the value obtained from the value of $N=66$ corrected
by $\Delta T_0$ and 2-body finite size corrections $\Delta FSE$ obtained
from interpolating
the structure factor as given in the tables.
The result for $N=66$ shown by the yellow square labeled with $\theta_\infty$ 
has been obtained with exact twist averaging\cite{fse1}.
}
\label{ExtrapolE0_N_rs70}
\end{figure}
\begin{figure}[h]
\includegraphics[width=8cm]{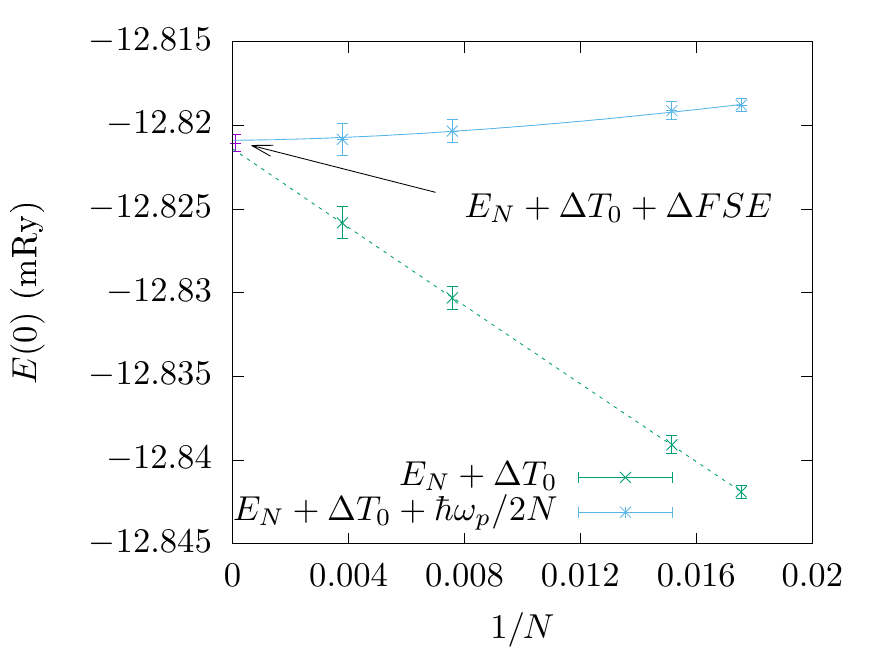}
\caption{
Numerical extrapolation of twist-averaged energies of 
BF VMC wave functions with analytical Jastrow  and backflow potential
for $N=57$, $N=66$, $132$, and $264$ electrons at $r_s=120$ and zero polarization
corrected by the ideal gas kinetic energy corrections, $E_N+\Delta T_0$, 
together with
the energy values including the analytical leading order 1/N (Plasmon) correction,
$E_N+\Delta T_0 + \hbar \omega_p/2N$ with
$\hbar \omega_p/2=\sqrt{3/r_s^3}$ Ry.
The lines are based on fitting linear and $N^{-5/3}$ corrections for $E_N+\Delta T_0$
and only $N^{-5/3}$ corrections for $E_N+\Delta T_0 + \hbar \omega_p/2N$.
At the origin we show the value obtained from the value of $N=66$ corrected
by $\Delta T_0$ and 2-body finite size corrections $\Delta FSE$ obtained
from interpolating
the structure factor as given in the tables.
}
\label{ExtrapolE0_N_BF_rs120}
\end{figure}
Leading order two-body corrections to the total energy are due to collective plasmon excitations
\cite{fse1,fse_rpa}. They only depend on the total density and are
not sensitive to spin polarization. Therefore,
the polarization energy is expected to approach the thermodynamic
limit more rapidly  than the total energy.
Figure~\ref{Extrapol_N_SJBF} shows the exchange-correlation part of the polarization energies
for our three densities from calculations with $N=66$ and $N=264$: systematic finite size bias
is not visible within the accuracy.

In order to check the relevance of quasirandom fluctuations \cite{Ruggeri18} we have performed
calculations for $N=57$ having a closed shell for the fully polarized state,
in contrast to $N=66$ with closed shell at the Gamma point for the unpolarized state.
From Fig.~\ref{Extrapol_N_SJBF}, we do not see any systematic bias within our uncertainty.
Furthermore, we have checked the energy differences between
the fully polarized and the unpolarized state at $r_s=70$, yielding
$E(\zeta=1)-E(\zeta=0)=42.5 (7) \mu$Ry  for $N=66$, compared to
$43.7(6) \mu$Ry for $N=57$, 
$41.4(9) \mu$ Ry for $N=132$, and $=41.8(15) \mu $Ry for $N=264$, all compatible
within the error. 
Note that using canonical twist average, for $N=57$ we are forced to consider slightly
different polarizations than for $N=66$ or 264, apart from the case of $\zeta=1$.
For $N=57$, the value of $E(\zeta=0)$
is estimated combining the calculated energy at $\zeta=1/57$,
the polynomial fit of the exchange correlation energy, and the
ideal gas polarization energy.

We have also compared the polarization energy
at $r_s=120$ between systems with $N=66$ and $N=132$ electrons 
using analytical BF wave function \cite{BF03} within VMC
(Figure~\ref{Extrapol_N_SJBF}),
yielding again no visible bias.

In Figure~\ref{ExtrapolE0_N_rs70} we illustrate the size effects of the total energy at zero 
polarization at $r_s=70$. Our finite size corrected energies are generally within 1 $\mu$Ry
of the numerical extrapolations. 
We take this value as an estimate of the systematic bias
in the extrapolation to the thermodynamic limit.
As another example, the
numerical extrapolation of the VMC data with analytical backflow wave functions \cite{BF03}
with $N=57$, $66$, $132$, and $264$ electrons at $r_s=120$,
shown in Fig.~\ref{ExtrapolE0_N_BF_rs120}, agrees with
the finite size corrected $N=66$ data within the uncertainty. 
The numerical extrapolations of Figs.~\ref{ExtrapolE0_N_rs70}
and \ref{ExtrapolE0_N_BF_rs120} are not sensitive to the precise power law
assumed for the next-to-leading-order term. In the figures, we show the extrapolation based on
 a $N^{-5/3}$ behavior, as resulting from a regular expansion of the
structure factor at the origin.

In the result for the Wigner crystallization density, $r_s=113(2)$, the uncertainty
includes both the statistical error on the Monte Carlo data and our estimate
of the systematic bias in the size corrections.

%\newpage

\subsection*{Tables of QMC results}

Tables \ref{table_rs70}, \ref{table_rs100}, and \ref{table_rs120} list 
all the finite--size energies 
and variances for different wave functions, densities and polarizations as
specified in the main text, as well as zero--variance extrapolations and 
finite--size corrections.

\begin{table}
\begin{tabular}{|c|c|c c c c c |}
\hline
$\zeta$ & $\Psi$ & $\sigma^2/\sigma_0^2$ & VMC & DMC & $\Delta T_0$  & $\Delta FSE$ \\
\hline
\multirow{8}{*}{0.00} 
  & SJ  & 1.00000(326) & -21.17385(39) & -21.31364(28) & -0.00106 & 0.04373 \\ \cline{2-7}
  & BF0 & 0.32706(91)  & -21.31461(21) & -21.35302(12) & -0.00106 & 0.04079 \\ \cline{2-7}
  & BF1 & 0.15819(51)  & -21.34883(12) & -21.36427(07) & -0.00106 & 0.04097 \\ \cline{2-7}
  & BF2 & 0.07838(25)  & -21.36009(08) & -21.36794(08) & -0.00106 & 0.04117 \\ \cline{2-7}
  & BF3 & 0.07079(28)  & -21.36121(08) & -21.36841(03) & -0.00106 & 0.04180 \\ \cline{2-7}
  & BF4 & --           & --            &--             & --       & --      \\ \cline{2-7}
  & ext full   &       & -21.37267(160)& -21.37238(56) &          &         \\ \cline{2-7}
  & ext w/o SJ &       & -21.36843(48) & -21.37090(26) &          &         \\ \cline{1-7}
\multirow{8}{*}{0.42} 
  & SJ  & 0.96797(304) & -21.18668(40) & -21.31704(29) & -0.00053 & 0.04373 \\ \cline{2-7}
  & BF0 & 0.30780(89)  & -21.31742(19) & -21.35256(09) & -0.00053 & 0.04079 \\ \cline{2-7}
  & BF1 & 0.13848(48)  & -21.34880(12) & -21.36240(06) & -0.00053 & 0.04097 \\ \cline{2-7}
  & BF2 & 0.08538(28)  & -21.35649(09) & -21.36481(08) & -0.00053 & 0.04117 \\ \cline{2-7}
  & BF3 & 0.07069(30)  & -21.35860(08) & -21.36555(04) & -0.00053 & 0.04180 \\ \cline{2-7}
  & BF4 & --           & --            &--             & --       & --      \\ \cline{2-7}
  & ext full   &       & -21.37013(102)& -21.36941(38) &          &         \\ \cline{2-7}
  & ext w/o SJ &       & -21.36710(32) & -21.36832(19) &          &         \\ \cline{1-7}
\multirow{8}{*}{0.61} 
  & SJ  & 0.91064(303) & -21.19929(46) &-21.32103(23)  & -0.00066 & 0.04373 \\ \cline{2-7}
  & BF0 & 0.28858(86)  & -21.32023(18) &-21.35184(07)  & -0.00066 & 0.04079 \\ \cline{2-7}
  & BF1 & 0.13152(42)  & -21.34817(10) &-21.36046(06)  & -0.00066 & 0.04097 \\ \cline{2-7}
  & BF2 & 0.08425(28)  & -21.35481(09) &-21.36249(04)  & -0.00066 & 0.04117 \\ \cline{2-7}
  & BF3 & 0.06870(56)  & -21.35706(09) &-21.36316(03)  & -0.00066 & 0.04180 \\ \cline{2-7}
  & BF4 & --           & --            &--             & --       & --      \\ \cline{2-7}
  & ext full   &       & -21.36794(97) &-21.36678(32)  &          &         \\ \cline{2-7}
  & ext w/o SJ &       & -21.36515(49) &-21.36563(13)  &          &         \\ \cline{1-7}
\multirow{8}{*}{0.79} 
  & SJ  & 0.81174(275) & -21.21608(37) & -21.32436(25) & -0.00074 & 0.04373 \\ \cline{2-7}
  & BF0 & 0.26009(70)  & -21.32214(17) & -21.34939(14) & -0.00074 & 0.04079 \\ \cline{2-7}
  & BF1 & 0.12528(34)  & -21.34535(09) & -21.35628(06) & -0.00074 & 0.04097 \\ \cline{2-7}
  & BF2 & 0.08368(27)  & -21.35111(07) & -21.35773(05) & -0.00074 & 0.04117 \\ \cline{2-7}
  & BF3 & 0.06924(48)  & -21.35314(07) & -21.35856(03) & -0.00074 & 0.04180 \\ \cline{2-7}
  & BF4 & --           & --            &--             & --       & --      \\ \cline{2-7}
  & ext full   &       & -21.36358(80) & -21.36170(40) &          &         \\ \cline{2-7}
  & ext w/o SJ &       & -21.36116(44) & -21.36097(96) &          &         \\ \cline{1-7}
\multirow{8}{*}{1.00} 
  & SJ  & 0.63694(203) & -21.23829(34) & -21.32189(21) & -0.00066 & 0.04373 \\ \cline{2-7}
  & BF0 & 0.21696(58)  & -21.31933(16) & -21.34051(07) & -0.00066 & 0.04079 \\ \cline{2-7}
  & BF1 & 0.10935(31)  & -21.33719(09) & -21.34574(04) & -0.00066 & 0.04097 \\ \cline{2-7}
  & BF2 & 0.07083(24)  & -21.34198(08) & -21.34700(05) & -0.00066 & 0.04117 \\ \cline{2-7}
  & BF3 & 0.05861(16)  & -21.34384(07) & -21.34774(05) & -0.00066 & 0.04180 \\ \cline{2-7}
  & BF4 & --           & --            &--             & --       & --      \\ \cline{2-7}
  & ext full   &       & -21.35177(67) & -21.35030(40) &          &         \\ \cline{2-7}
  & ext w/o SJ &       & -21.35006(112)& -21.34945(77) &          &         \\
\hline

\end{tabular}
\caption{
Energy per particle (in mRy) at various spin polarizations $\zeta$ for $r_s=70$ 
from VMC and DMC simulations of 66 electrons in twist-averaged boundary 
conditions using different wave functions, SJ and BF$k$; 
variance of the local energy relative to that of the SJ wave function at $\zeta=0$, $\sigma^2/\sigma_0^2$;
zero--variance extrapolation of the VMC and DMC energies with or without the SJ result
(``ext full'' and ``ext w/o SJ'', obtained using gnuplot and taking into account  
the standard deviations of both energy and variance);
finite size errors for the discretization of $k$-space through the non--interacting shell effect, $\Delta T_0$, Eq.(22) of Ref.~\onlinecite{fse2},
and through integrals involving the static structure factor, $\Delta FSE$, Eq.(30), (35), and (38) of Ref.~\onlinecite{fse2}. In the calculation of $\Delta FSE$ we use the RPA two--body pseudopotential
and the analytic backflow; we also average over the polarizations because there is not enough statistical precision 
to detect a polarization dependence.
        }
\label{table_rs70}
\end{table}

\begin{table}
\begin{tabular}{|c|c|c c c c c |}
\hline
$\zeta$ & $\Psi$ & $\sigma^2/\sigma_0^2$ & VMC & DMC & $\Delta T_0$  & $\Delta FSE$ \\
\hline
\multirow{8}{*}{0.00} 
  & SJ  & 1.00000(232) & -15.25228(13) & -15.35243(28) & -0.00052 & 0.02543 \\ \cline{2-7}
  & BF0 & 0.34206(80)  & -15.34443(13) & -15.37588(09) & -0.00052 & 0.02382 \\ \cline{2-7}
  & BF1 & 0.15445(43)  & -15.37101(08) & -15.38345(07) & -0.00052 & 0.02413 \\ \cline{2-7}
  & BF2 & 0.09342(23)  & -15.37899(06) & -15.38588(03) & -0.00052 & 0.02414 \\ \cline{2-7}
  & BF3 & 0.07226(17)  & -15.38112(05) & -15.38661(03) & -0.00052 & 0.02441 \\ \cline{2-7}
  & BF4 & 0.06484(17)  & -15.38189(04) & -15.38683(04) & -0.00052 & 0.02468 \\ \cline{2-7}
  & ext full   &       & -15.39050(57) & -15.38953(13) &          &         \\ \cline{2-7}
  & ext w/o SJ &       & -15.38886(59) & -15.38914(17) &          &         \\ \cline{1-7}
\multirow{8}{*}{0.18}
  & SJ  & 0.98798(413) & -15.25354(34) & -15.35302(08) & -0.00039 & 0.02543 \\ \cline{2-7}
  & BF0 & 0.33780(130) & -15.34508(18) & -15.37524(02) & -0.00039 & 0.02382 \\ \cline{2-7}
  & BF1 & 0.15244(46)  & -15.37101(11) & -15.38289(02) & -0.00039 & 0.02413 \\ \cline{2-7}
  & BF2 & 0.09258(33)  & -15.37890(08) & -15.38536(02) & -0.00039 & 0.02414 \\ \cline{2-7}
  & BF3 & 0.07234(31)  & -15.38086(07) & -15.38604(01) & -0.00039 & 0.02441 \\ \cline{2-7}
  & BF4 & 0.06409(27)  & -15.38161(06) & -15.38636(02) & -0.00039 & 0.02468 \\ \cline{2-7}
  & ext full   &       & -15.39001(63) & -15.38910(13) &          &         \\ \cline{2-7}
  & ext w/o SJ &       & -15.38847(81) & -15.38876(15) &          &         \\ \cline{1-7}
\multirow{8}{*}{0.42} 
  & SJ  & 0.91178(131) & -15.26052(11) & -15.35652(30) & -0.00026 & 0.02543 \\ \cline{2-7}
  & BF0 & 0.32012(77)  & -15.34784(12) & -15.37672(08) & -0.00026 & 0.02382 \\ \cline{2-7}
  & BF1 & 0.14920(24)  & -15.37188(06) & -15.38343(06) & -0.00026 & 0.02413 \\ \cline{2-7}
  & BF2 & 0.09047(25)  & -15.37929(06) & -15.38558(03) & -0.00026 & 0.02414 \\ \cline{2-7}
  & BF3 & --           & --            & --            & --       & --      \\ \cline{2-7}
  & BF4 & 0.06415(18)  & -15.38196(05) & -15.38645(03) & -0.00026 & 0.02468 \\ \cline{2-7}
  & ext full   &       & -15.39015(69) & -15.38899(16) &          &         \\ \cline{2-7}
  & ext w/o SJ &       & -15.38876(64) & -15.38857(09) &          &         \\ \cline{1-7}
\multirow{8}{*}{0.61} 
  & SJ  & 0.89692(241) & -15.26984(24) &-15.35928(33)  & -0.00032 & 0.02543 \\ \cline{2-7}
  & BF0 & 0.30266(85)  & -15.35090(12) &-15.37749(10)  & -0.00032 & 0.02382 \\ \cline{2-7}
  & BF1 & 0.14164(36)  & -15.37289(07) &-15.38338(06)  & -0.00032 & 0.02413 \\ \cline{2-7}
  & BF2 & 0.08791(24)  & -15.37962(05) &-15.38536(04)  & -0.00032 & 0.02414 \\ \cline{2-7}
  & BF3 & 0.07001(65)  & -15.38172(15) &-15.38582(03)  & -0.00032 & 0.02441 \\ \cline{2-7}
  & BF4 & 0.06261(18)  & -15.38209(04) &-15.38611(03)  & -0.00032 & 0.02468 \\ \cline{2-7}
  & ext full   &       & -15.39009(58) &-15.38846(13)  &          &         \\ \cline{2-7}
  & ext w/o SJ &       & -15.38857(62) &-15.38811(21)  &          &         \\ \cline{1-7}
\multirow{8}{*}{0.79} 
  & SJ  & 0.81249(244) & -15.28176(22) & -15.36306(23) & -0.00036 & 0.02543 \\ \cline{2-7}
  & BF0 & 0.27848(73)  & -15.35429(12) & -15.37792(09) & -0.00036 & 0.02382 \\ \cline{2-7}
  & BF1 & 0.13232(37)  & -15.37342(06) & -15.38282(03) & -0.00036 & 0.02413 \\ \cline{2-7}
  & BF2 & 0.08441(22)  & -15.37923(05) & -15.38435(02) & -0.00036 & 0.02414 \\ \cline{2-7}
  & BF3 & 0.06808(16)  & -15.38081(05) & -15.38484(02) & -0.00036 & 0.02441 \\ \cline{2-7}
  & BF4 & 0.06179(20)  & -15.38135(04) & -15.38499(02) & -0.00036 & 0.02468 \\ \cline{2-7}
  & ext full   &       & -15.38877(44) & -15.38706(10) &          &         \\ \cline{2-7}
  & ext w/o SJ &       & -15.38758(52) & -15.38679(09) &          &         \\ \cline{1-7}
\multirow{8}{*}{1.00} 
  & SJ  & 0.67302(100) & -15.29788(10) & -15.36486(22) & -0.00032 & 0.02543 \\ \cline{2-7}
  & BF0 & 0.24224(67)  & -15.35647(10) & -15.37570(08) & -0.00032 & 0.02382 \\ \cline{2-7}
  & BF1 & 0.11261(29)  & -15.37226(07) & -15.37986(05) & -0.00032 & 0.02413 \\ \cline{2-7}
  & BF2 & 0.07619(22)  & -15.37678(05) & -15.38098(04) & -0.00032 & 0.02414 \\ \cline{2-7}
  & BF3 & 0.06177(20)  & -15.37821(03) & -15.38141(02) & -0.00032 & 0.02441 \\ \cline{2-7}
  & BF4 & 0.05502(14)  & -15.37866(04) & -15.38163(03) & -0.00032 & 0.02468 \\ \cline{2-7}
  & ext full   &       & -15.38499(35) & -15.38347(06) &          &         \\ \cline{2-7}
  & ext w/o SJ &       & -15.38435(81) & -15.38325(03) &          &         \\ 
\hline

\end{tabular}
\caption{
Same as Table \ref{table_rs70} for $r_s=100.$
        }
\label{table_rs100}
\end{table}

\begin{table}
\begin{tabular}{|c|c|c c c c c |}
\hline
$\zeta$ & $\Psi$ & $\sigma^2/\sigma_0^2$ & VMC & DMC & $\Delta T_0$  & $\Delta FSE$ \\
\hline
\multirow{8}{*}{0.00} 
  & SJ  & 1.0000(30)   & -12.87447(19) & -12.95868(20) & -0.00036 & 0.01930 \\ \cline{2-7}
  & BF0 & 0.3557(10)   & -12.94860(12) & -12.97655(07) & -0.00036 & 0.01803 \\ \cline{2-7}
  & BF1 & 0.1728(04)   & -12.97104(05) & -12.98269(08) & -0.00036 & 0.01883 \\ \cline{2-7}
  & BF2 & 0.1026(03)   & -12.97875(04) & -12.98476(03) & -0.00036 & 0.01876 \\ \cline{2-7}
  & BF3 & 0.0804(02)   & -12.98078(04) & -12.98557(03) & -0.00036 & 0.01886 \\ \cline{2-7}
  & BF4 & 0.0716(02)   & -12.98137(04) & -12.98572(02) & -0.00036 & 0.01893 \\ \cline{2-7}
  & ext full   &       & -12.98943(65) & -12.98819(12) &          &         \\ \cline{2-7}
  & ext w/o SJ &       & -12.98782(59) & -12.98786(27) &          &         \\ \cline{1-7}
\multirow{8}{*}{0.18}
  & SJ  & 0.97644(451) & -12.87615(25) & -12.96060(34) & -0.00027 & 0.01930 \\ \cline{2-7}
  & BF0 & 0.35345(134) & -12.94900(13) & -12.97625(32) & -0.00027 & 0.01803 \\ \cline{2-7}
  & BF1 & 0.17149(62)  & -12.97128(09) & -12.98218(25) & -0.00027 & 0.01883 \\ \cline{2-7}
  & BF2 & 0.10206(36)  & -12.97862(07) & -12.98456(51) & -0.00027 & 0.01876 \\ \cline{2-7}
  & BF3 & 0.07986(30)  & -12.98060(06) & -12.98509(34) & -0.00027 & 0.01886 \\ \cline{2-7}
  & BF4 & 0.07087(28)  & -12.98124(06) & -12.98538(23) & -0.00027 & 0.01893 \\ \cline{2-7}
  & ext full   &       & -12.98900(68) & -12.98786(14) &          &         \\ \cline{2-7}
  & ext w/o SJ &       & -12.98720(43) & -12.98758(12) &          &         \\ \cline{1-7}
\multirow{8}{*}{0.42} 
  & SJ  & 0.96287(299) & -12.88101(24) & -12.96112(18) & -0.00018 & 0.01930 \\ \cline{2-7}
  & BF0 & 0.33762(89)  & -12.95121(10) & -12.97750(06) & -0.00018 & 0.01803 \\ \cline{2-7}
  & BF1 & 0.16520(45)  & -12.97204(06) & -12.98279(03) & -0.00018 & 0.01883 \\ \cline{2-7}
  & BF2 & 0.09907(26)  & -12.97913(04) & -12.98477(03) & -0.00018 & 0.01876 \\ \cline{2-7}
  & BF3 & 0.07704(19)  & -12.98108(04) & -12.98537(03) & -0.00018 & 0.01886 \\ \cline{2-7}
  & BF4 & 0.06876(19)  & -12.98179(03) & -12.98564(03) & -0.00018 & 0.01893 \\ \cline{2-7}
  & ext full   &       & -12.98943(57) & -12.98780(07) &          &         \\ \cline{2-7}
  & ext w/o SJ &       & -12.98777(37) & -12.98760(05) &          &         \\ \cline{1-7}
\multirow{8}{*}{0.61} 
  & SJ  & 0.91729(294) & -12.88814(19) &-12.96447(21)  & -0.00022 & 0.01930 \\ \cline{2-7}
  & BF0 & 0.32041(99)  & -12.95415(10) &-12.97834(06)  & -0.00022 & 0.01803 \\ \cline{2-7}
  & BF1 & 0.15444(40)  & -12.97320(05) &-12.98313(04)  & -0.00022 & 0.01883 \\ \cline{2-7}
  & BF2 & 0.09523(23)  & -12.97963(05) &-12.98488(03)  & -0.00022 & 0.01876 \\ \cline{2-7}
  & BF3 & 0.07506(18)  & -12.98150(04) &-12.98546(02)  & -0.00022 & 0.01886 \\ \cline{2-7}
  & BF4 & 0.06584(20)  & -12.98211(04) &-12.98562(02)  & -0.00022 & 0.01893 \\ \cline{2-7}
  & ext full   &       & -12.98939(48) &-12.98768(09)  &          &         \\ \cline{2-7}
  & ext w/o SJ &       & -12.98822(63) &-12.98748(17)  &          &         \\ \cline{1-7}
\multirow{8}{*}{0.79} 
  & SJ  & 0.83963(256) & -12.89765(20) & -12.96749(13) & -0.00025 & 0.01930 \\ \cline{2-7}
  & BF0 & 0.29635(74)  & -12.95744(10) & -12.97902(11) & -0.00025 & 0.01803 \\ \cline{2-7}
  & BF1 & 0.14659(39)  & -12.97407(07) & -12.98321(03) & -0.00025 & 0.01883 \\ \cline{2-7}
  & BF2 & 0.09018(23)  & -12.97979(04) & -12.98461(03) & -0.00025 & 0.01876 \\ \cline{2-7}
  & BF3 & 0.07189(18)  & -12.98147(03) & -12.98508(02) & -0.00025 & 0.01886 \\ \cline{2-7}
  & BF4 & 0.06351(18)  & -12.98203(04) & -12.98522(02) & -0.00025 & 0.01893 \\ \cline{2-7}
  & ext full   &       & -12.98867(39) & -12.98689(11) &          &         \\ \cline{2-7}
  & ext w/o SJ &       & -12.98751(44) & -12.98663(08) &          &         \\ \cline{1-7}
\multirow{8}{*}{1.00} 
  & SJ  & 0.71979(221) & -12.91080(17) & -12.96998(18) & -0.00022 & 0.01930 \\ \cline{2-7}
  & BF0 & 0.26296(59)  & -12.96006(09) & -12.97809(05) & -0.00022 & 0.01803 \\ \cline{2-7}
  & BF1 & 0.12702(33)  & -12.97426(05) & -12.98178(04) & -0.00022 & 0.01883 \\ \cline{2-7}
  & BF2 & 0.08318(39)  & -12.97891(05) & -12.98289(02) & -0.00022 & 0.01876 \\ \cline{2-7}
  & BF3 & 0.06682(18)  & -12.98021(04) & -12.98324(02) & -0.00022 & 0.01886 \\ \cline{2-7}
  & BF4 & 0.06109(35)  & -12.98067(03) & -12.98338(02) & -0.00022 & 0.01893 \\ \cline{2-7}
  & ext full   &       & -12.98687(25) & -12.98516(12) &          &         \\ \cline{2-7}
  & ext w/o SJ &       & -12.98632(53) & -12.98470(07) &          &         \\ 
\hline

\end{tabular}
\caption{
Same as Table \ref{table_rs70} for $r_s=120.$
        }
\label{table_rs120}
\end{table}

%\newpage
%


\begin{thebibliography}{99}
\bibitem{Bloch1929} F. Bloch, Zeitschrift f. Physik {\bf 57}, 545 (1929).
\bibitem{Wigner1934} E. P. Wigner, Phys. Rev. 46, 1002 (1934).
\bibitem{Wigner1938} E. P. Wigner, Trans. Farad. Soc. {\bf 34}, 678 (1938).
\bibitem{Stoner1938} E.C. Stoner, Proc. R. Soc. Lond. A {\bf 165}, 372 (1938).
%\bibitem{Schubin34} S. Schubin and S. Wonsowsky, Proc. Roy. Soc. A {\bf 145}, 159 (1934).
\bibitem{Ceperley78} D. Ceperley, Phys. Rev. B {\bf 18}, 3126 (1978).
\bibitem{CeperleyAlder} D. M. Ceperley and B. J. Alder, Phys. Rev. Lett. {\bf 45}, 566 (1980).
\bibitem{CAbis} D. M. Ceperley and B. J. Alder, Journal de Physique C{\bf 7}, 295 (1980).
\bibitem{ACP} B.J. Alder, D.M. Ceperley, and E.L. Pollock, Int. J. Quant. Chem. {\bf 16}, 49 (1982).
\bibitem{Ortiz1999} G. Ortiz, M. Harris, and P. Ballone,
Phys. Rev. Lett. {\bf 82}, 5317 (1999).
\bibitem{Zong2002} F. H. Zong, C. Lin, and D. M. Ceperley,
Phys. Rev. E {\bf 66}, 036703 (2002).
\bibitem{Drummond04} N. D. Drummond, Z. Radnai, J. R. Trail, M. D. Towler, and R. J. Needs,
Phys. Rev. B {\bf 69}, 085116 (2004).
\bibitem{Young99} D.P. Young, D. Hall, M.E. Torelli, Z. Fisk, J.L. Sarrao, J.D. Thompson, H.-R. Ott, S.B. Oseroff, R.G. Goodrich, and R. Zysler, Nature (London) {\bf 397}, 412 (1999).
\bibitem{Ichimaru} S. Ichimaru, Phys. Rev. Lett. {\bf 84}, 1842 (2000).
\bibitem{OritizC} G. Ortiz, M. Harris, and P. Ballone, Phys. Rev. Lett. {\bf 84}, 1843 (2000).
\bibitem{Ketterle} G.-B. Jo1, Y.-R. Lee, J.-H. Choi, C. A. Christensen, T. H. Kim, J. H. Thywissen, D. E. Pritchard, W. Ketterle, Science {\bf 325}, 1521 (2009).
\bibitem{Firenze} G. Valtolina, F. Scazza, A. Amico, A. Burchianti, A. Recati, T. Enss, M. Inguscio, M. Zaccanti, and G. Roati, Nature Physics {\bf 13}, 704 (2017).
\bibitem{bfiter1} M. Taddei, M. Ruggeri, S. Moroni, and M. Holzmann,
Phys. Rev. B {\bf 91}, 115106 (2015).
\bibitem{bfiter2} M. Ruggeri, S. Moroni, and M. Holzmann,
Phys. Rev. Lett. {\bf 120}, 205302 (2018).
\bibitem{Wagner16} L.K. Wagner and D.M. Ceperley,  Rep. Prog. Phys. {\bf 79}, 094501 (2016).
\bibitem{fpa} G. Ortiz, D.M. Ceperley, and R.M. Martin, Phys. Rev. Lett. {\bf 71}, 2777 (1993).
\bibitem{Lin2001} C. Lin, F.-H. Zong, and D. M. Ceperley, Phys. Rev. E {\bf 64}, 016702
(2001).
\bibitem{fse1} S. Chiesa, D.M. Ceperley, R.M. Martin, and M. Holzmann,
Phys. Rev. Lett. {\bf 97}, 076404 (2006).
\bibitem{fse2} M. Holzmann, R.C. Clay III, M. A. Morales, N.M. Tubman, D. M. Ceperley, and C. Pierleoni,
Phys. Rev. B {\bf 94}, 035126 (2016).
\bibitem{drummond08} N.D. Drummond, R.J. Needs, A. Sorouri, and W.M.C. Foulkes, Phys. Rev. B {\bf 78}, 125106 (2008).
%\bibitem{Tanatar} B. Tanatar and D. M. Ceperley, Phys. Rev. B {\bf 39}, 5005 (1989).
\bibitem{Rapisarda} F. Rapisarda and G. Senatore, Aust. J. Phys. {\bf 49}, 161 (1996).
\bibitem{Varsano} D. Varsano, S. Moroni and G. Senatore, Europhys. Lett. {\bf 53}, 348 (2001).
\bibitem{Attac} C. Attaccalite, S. Moroni, P. Gori-Giorgi, and G. B. Bachelet, Phys. Rev. Lett. 88, 256601 (2002).
\bibitem{Drummond09} N. D. Drummond and R. J. Needs
Phys. Rev. Lett. {\bf 102}, 126402 (2009).
\bibitem{vignale} G. F. Giuliani and G. Vignale, Quantum Theory of the Electron
Liquid (Cambridge University Press, Cambridge, 2005).
\bibitem{book} R.M. Martin, L. Reining, and D. M. Ceperley, {\cal Interacting Electrons}, Cambridge  University Press, Cambridge (2016).
\bibitem{Ewald} P.P. Ewald, Ann. Phys. {\bf 64}, 253 (1921).
\bibitem{Natoli} V. Natoli and D.M. Ceperley, J. Comput. Phys. {\bf 117}, 171 (1995).
\bibitem{supp} See Supplemental Material for 
details of the simulations, interpolations of the polarization energy, 
finite--size corrections, and tables with all QMC data and their zero--variance 
extrapolations. 
\bibitem{BF03} M. Holzmann, D.M. Ceperley, C. Pierleoni, and K. Esler,
Phys. Rev. E {\bf 68}, 046707 (2003).
\bibitem{fse_rpa} M. Holzmann, B. Bernu, and D. M. Ceperley,
J. Phys.: Conf. Ser. {\bf 321} 012020 (2011).
\bibitem{Ruggeri18} M. Ruggeri, P. L{\'o}pez R{\'i}os, and A. Alavi, Phys. Rev. B {\bf 98}, 161105(R) (2018).
\bibitem{Spink13} G.G. Spink, R.J. Needs, and N.D. Drummond, Phys. Rev. B {\bf 88}, 085121 (2013).
\bibitem{Nava12} M. Nava, A. Motta, D. E. Galli, E. Vitali, and S. Moroni,
Phys. Rev. B {\bf 85}, 184401 (2012).
\bibitem{he} M. Holzmann, B. Bernu, and D.M. Ceperley,
Phys. Rev. B {\bf 74}, 104510 (2006).
\bibitem{dipol} T. Comparin, R. Bombin, M. Holzmann, F. Mazzanti, J. Boronat, and S. Giorgini,
Phys. Rev. A {\bf 99}, 043609 (2019).
\bibitem{Overhauser} A. W. Overhauser, Phys. Rev. Lett. {\bf 4}, 462 (1960); Phys. Rev.
{\bf 128}, 1437 (1962).
\bibitem{HF} F. Delyon, B. Bernu, L. Baguet, and M. Holzmann,
Phys. Rev. B {\bf 92}, 235124 (2015).
\bibitem{Lewin} D. Gontier, C. Hainzl, and M. Lewin,
Phys. Rev. A {\bf 99}, 052501 (2019).
\end{thebibliography}
\end{document}